\documentclass[twocolumn]{revtex4-1}
\usepackage{graphicx,hyperref,amsmath}
\usepackage[matrix,frame,arrow]{xy}
\usepackage{color}
\usepackage[vcentermath]{youngtab}
\newcommand{\bra}[1]{\langle#1|}
\newcommand{\ket}[1]{|#1\rangle}

\newcommand{\qw}[1][-1]{\ar @{-} [0,#1]}
\newcommand{\qwx}[1][-1]{\ar @{-} [#1,0]}

\newcommand{\gate}[1]{*{\xy *+<.6em>{#1};p\save+LU;+RU **\dir{-}\restore\save+RU;+RD **\dir{-}\restore\save+RD;+LD **\dir{-}\restore\POS+LD;+LU **\dir{-}\endxy} \qw}

\newcommand{\control}{*-=-{\bullet}}

\newcommand{\ctrl}[1]{\control \qwx[#1] \qw}
\newcommand{\rstick}[1]{*!L!<-.5em,0em>=<0em>{#1}}
\newcommand{\lstick}[1]{*!R!<.5em,0em>=<0em>{#1}}
\newcommand{\ustick}[1]{*!D!<0em,-.5em>=<0em>{#1}}

\newcommand{\Qcircuit}{\xymatrix @*=<0em>}
\begin{document}
\date{\today}
\title{Spin-free quantum computational simulations and symmetry adapted states}
\author{James Daniel Whitfield} 
\email[email: ]{james.whitfield@univie.ac.at}
\affiliation{Vienna Center for Quantum Science and Technology}
\affiliation{University of Vienna}

\begin{abstract}
The ideas of digital simulation of quantum systems using a quantum computer parallel the original ideas of numerical simulation using a classical computer.  In order for quantum computational simulations to advance  to a competitive point, many techniques from classical simulations must be imported into the quantum domain.  In this article, we consider the applications of symmetry in the context of quantum simulation.  Building upon well established machinery, we propose a form of first quantized simulation that only requires the spatial part of the wave function, thereby allowing spin-free quantum computational simulations.  We go further and discuss the preparation of $N$-body states with specified symmetries based on projection techniques. We consider two simple examples, molecular hydrogen and cyclopropenyl cation, to illustrate the ideas.  While the methods here represent adaptations of known quantum algorithms, they are the first to explicitly deal with preparing $N$-body symmetry-adapted states.
\end{abstract}
\maketitle

In 1929, Dirac noted~\cite{Dirac29} ``The underlying physical laws necessary for the mathematical theory of a large part of physics and the whole of chemistry are thus completely known, and the difficulty is only that the exact application of these laws leads to equations much to complicated to be soluble." In the 84 years since this declaration, there has been a constant push against this computational complexity with increasingly sophisticated computational software and hardware technologies. In this article, we will pay particular attention to the methods being developed around the suggestion of Feynman~\cite{Feynman82} to use quantum resources to simulate quantum systems~\cite{Whitfield11,Veis10,Yung12,*Kassal11,*Brown10,*Buluta09}. 

In quantum computational simulation, solving the Schr\"odinger equation is predicated on the ability of quantum computers to efficiently evolve a quantum state.  The quantum phase estimation algorithm~\cite{Kitaev95,*Cleve98} is, at its essence, a Fourier transform of the auto-correlation function $\langle \psi(0)\ket{\psi(t)}$ in order to obtain a spectrum containing the eigen-frequencies \cite{Whitfield11,Wang12}.  The frequencies can be converted into energies provided one avoids aliasing the high and low energies in the conversion process.  Such Fourier transform methods require that the input state have non-trivial overlap with the eigenstate of interest.  The overlap of a random state with a specific state is expected to be inversely proportional to the dimension of the vector space. For systems of composite particles, it is intuitive and correct, to believe that preparing a state with sufficient overlap to an eigenstate may be as difficult as solving a quantum NP-hard problem~\cite{Kitaev02,Whitfield13}. In this general setting, by projecting into an targeted subspace, we increase the overlap between a random state and the ground state which facilitates the QMA-hard task of quantum state preparation. 

The methods presented here are aimed at closing the gap between the quantum and classical numerical simulations by properly accounting for symmetry in the quantum computational setting. Despite initial hesitation to adopt group theoretic methods~\cite{Slater29}, all modern numerical methods in quantum chemistry incorporate ideas developed based on group theory~\cite{Pauncz95,*Wigner59}.  Indeed, work on spin-free quantum chemistry has generate a large literature e.g. F. A. Matsen published a series of papers (`Spin-free quantum chemistry') spanning 15 years and some 20 papers beginning in 1965~\cite{Matsen65}.  The current paper is the first of its kind from the quantum simulation community.
 

The exploitation of symmetry in quantum simulation has been implicitly used but these have either been ad-hoc \cite{Lanyon10,*Du10}, classically prepared~\cite{Ortiz01,Wang08,Veis10} or only concerned with totally antisymmetric or symmetric states~\cite{Abrams97,Ward09,Kassal08,*Zalka98,*Wiesner96}.  The most closely related work to the present article, Ref.~\cite{Ward09}, includes a technical review of quantum state preparation and a discussion of preparing symmetry adapted single-electron orbitals but stopped short of considering $N$-body symmetry adapted states as done here.  In the context of quantum communication and quantum coding theory, symmetry-adapted states for the permutation and unitary groups were prepared using the quantum Schur-Weyl transform \cite{Bacon06,*Bacon07,Berg12}. 

In this article, we show, first, how to perform spin-free quantum simulations in first quantization by exploiting the permutation group and, second, how to project wave functions into the  irreducible representations (IR) corresponding to arbitrary symmetries. In general, the  symmetry group (group of constants of motion) of the Hamiltonian is defined by $\{A\;:\;[H,A]=0,\text{ }A^\dag=A, \text{ det}(A)\neq 0\}$.  This is a group since the identity commutes with the Hamiltonians and the basic commutator identity, 
\begin{equation}
	[gf,H]=g[f,H]+[g,H]f
	\label{eq:1}
\end{equation}
implies closure under multiplication. Assuming $g$ is a non-zero symmetry element of $H$ and substituting $f=g^{-1}$ into \eqref{eq:1}, we have $0=g[g^{-1},H]$ which implies that $g^{-1}$ also commutes with $H$. 
For non-relativistic quantum Hamiltonians, $S^2=S_x.S_x+S_y.S_y+S_z.S_z$ and $S_z=\sum_i s_z(i)$ are part of the symmetry group and for this reason spin eigenstates have been extensively considered in the physics and chemistry communities~\cite{Pauncz95,*Wigner59}.

We briefly recall the necessary facts about the symmetric group.  The symmetric (or permutation) group $S_N$ contains all permutations of $N$ items.  The IRs are labeled by Young tableaux or, equivalently, by a partition of $N$ objects, $[\lambda] = [\lambda_1, \lambda_2, ... , \lambda_k]$, with $\sum \lambda_i=N$ and $\lambda_i\geq\lambda_{i+1}\geq0$.  The numbers in the $[\lambda]$ vector indicate the length of the disjoint cycles in the permutation.  
The basis functions of each IR are obtained by the standard Young tableaux where the rows and columns of a Young's tableau are filled with numbers $1$ through $N$ such that the numbers increase along rows and down columns.  For example, 
\begin{equation}
	T_1=\young(12,34,5)
	\label{eq:tab}
\end{equation}
The projection into these states is done by translating a tableau with $n_r$ rows and $n_c$ columns  into the Young operator $E_i^{[\lambda]}=NP$ that symmetrizes the columns and antisymmetrizes the rows.  Here $N=\mathcal{A}_1\mathcal{A}_2...\mathcal{A}_{n_c}$ with $\mathcal{A}_i$ corresponding to the antisymmetrizer for the elements of the $i$-th row and $P=\mathcal{S}_1\mathcal{S}_2...\mathcal{S}_{n_r}$ with $\mathcal{S}_j$, the symmetrizer for the $j$-th row. 

A Schmidt decomposition for wave function, $\Psi(x_1,x_2,...,x_N)$, with $x_i=(r_i,s_i)$ is possible such that $\Psi=\sum F_i(r_1,...,r_N)\Theta_i(s_1,...,s_N)$. 
The functions $F_i$ and $\Theta_i$ are correlated so that the total state is completely symmetric (antisymmetric). This means that if $F_i$ transforms like the $i$-th basis function of IR $[\lambda]$ then,  for the total state to be symmetric (antisymmetric), the corresponding spin function must be the $i$-th basis function of the same (transposed) tableau. As an example, the transpose of \eqref{eq:tab} is $\tilde{T}_1=${\scriptsize$\young(135,24)$}.  Note that the partition $[\lambda]$ is constrained to have only two rows (there are only two spin functions $\alpha$ and $\beta$) and the difference in row lengths $\lambda_1-\lambda_2$ dictates the multiplicity of the spin state (singlet, doublet, etc.). 

Next, we transfer these ideas to make the necessary changes to the Abrams-Lloyd (AL) symmetrization/antisymmetrization algorithm~\cite{Abrams97} so that one can directly project into a spatial wave function with the desired symmetry.  This allows first quantized simulations to become \textit{spin-free} simulations.

The AL algorithm was proposed to construct fermionic and bosonic states for use in first quantized simulation. In the first quantization picture, one has $N$ registers of $M$-level systems (or $\log_2M$ qubits) as $\ket{\phi_i}\ket{\phi_j}...\ket{\phi_k}$ representing the Hartree product.  To perform the AL algorithm, prepare $N$ registers of $\log_2N$ qubits in state $\ket{B}=\sum_{b_1=1}^N\sum_{b_2=1}^{N-1}..\sum_{b_N=1}^1\ket{b_1b_2...b_N}/\sqrt{N!}$ which can be done in $O(N^2 \log M)$ operations \footnote{To ensure the classical algorithm is always reversible, one must include a second ancilla register and introduce some redundancy into the algorithm.}.  Then one uses a reversible classical algorithm: $(b_1b_2...b_N)\mapsto (b_1'b_2'...b_N')$ where $b_1'=b_1$ and $b_i'$ is the $b_i^{\textrm{th}}$ number not in $b_1'b_2'...b_{i-1}'$. The final step is to coherently sort  $\ket{B}$ (with appropriate phases for antisymmetrization) in parallel with the register of single particle wave functions.   Thus, 
\begin{eqnarray}
&&	\frac{1}{\sqrt{N!}}\ket{\phi_i\phi_j...\phi_k}\otimes\left(\sum_{b_1=1}^N\sum_{b_2=1}^{N-1}..\sum_{b_N=1}^1\ket{b_1b_2...b_N}\right)\phantom{sk}\\
&\mapsto&\frac{1}{\sqrt{N!}} \sum_{\pi\in S_N}\ket{\phi_i\phi_j...\phi_k}\otimes\ket{\pi(1...N)} \\
&\mapsto& \ket{\mathcal{S}(\phi_i\phi_j...\phi_k)}\otimes\ket{1...N}.
\end{eqnarray}

Symmetry of the wave functions on quantum computers has been discussed in completely symmetric or completely antisymmetric states \cite{Abrams97,Ward09}, but a simple extension of the AL algorithm can be used to prepare spin eigenstates.  Begin by selecting only the functions in the first row of a tableau and performing the symmetrization algorithm on these orbitals, then selecting the functions in the second row to be symmetrized and so on.  Then one continues by antisymmetrizing the columns in the same fashion.  As an example, consider $N=5$ with $S=1/2$ with tableaux $T_1$ as in \eqref{eq:tab}.   The algorithm proceeds via 
\begin{eqnarray}	
	&&\ket{\phi_i\phi_j\phi_k\phi_l\phi_m}\ket{12345}\\
	&\mapsto&\ket{\phi_i\phi_j\phi_k\phi_l\phi_m}\ket{\mathcal{A}(135)}\ket{\mathcal{A}(24)}\ket{\mathcal{S}(12)}\ket{\mathcal{S}(34)}\\
	&\mapsto	&\ket{E^{T} (\phi_i\phi_j\phi_k\phi_l\phi_m)}.
\end{eqnarray}
In the final step, the four $\ket{B}$ registers are unsorted from right to left.
The algorithms used to implement the Hamiltonian evolution in first quantization \cite{Kassal08,*Zalka98,*Wiesner96} only depend on the spatial coordinates and thus \textit{spin-free} simulations can be done with the properly prepared spatial states.

Now we turn to more general symmetry adaptation that is appropriate for other symmetries (e.g.~point groups or angular momentum) and for second quantization.  
Although orbitals that are variationally obtained such as the Hartree-Fock solution do not necessarily respect the symmetries of the Hamiltonian (known as L\"owdin's dilemma \cite{RMP63}),
the single electron molecular orbital basis can be symmetry adapted.  This can be done using the Wigner projection operators \cite{Pauncz95,*Wigner59}:
\begin{equation}
\mathbf{P}^{\Gamma}_{ij}=\frac{h_\Gamma}{h}\sum_{g\in G} D^{\Gamma}_{ij}(g)^* \;g
\end{equation}
where $\Gamma$ is the IR label, $D^{\Gamma}(g)$ is the irreducible $h_{\Gamma}\times h_{\Gamma}$ matrix representation of group member $g$ and $h$ is the number of elements in the group. On arbitrary functions, $F_0=P^\Gamma_{00}F$ and $F_i=P_{i0}F_0$ give a complete orthonormal basis for the irreducible representation (assuming that $F$ has overlap with all states).   If, as is often the case, the irreducible representations are not known, one can use, $P_\chi^\Gamma=\sum_i P^\Gamma_{ii}$, which only requires the characters (the trace of the representation matrix) of the IR of the group member. Tables of characters have been tabulated in many cases of interest such as point groups and are simple to calculate for certain groups such as the symmetric group.  

Implementing the Wigner projection operator can be done using a recently developed technique for probabilistic addition of unitary matrices~\cite{Childs12}, however this method is of limited utility in the current situation. The algorithm has the lowest success probability when the absolute value of the couplings are uniform as is often case when implementing the Wigner projection operators. Instead,
to obtain symmetry adapted $N$-body states, the phase estimation algorithm can be modified to perform projections into IRs \cite{Bacon06,*Bacon07}. 

First, we need the Fourier transform over an arbitrary group \cite{Moore04}, $G$,
\begin{equation}
	F_G=\frac{1}{\sqrt{h}}\sum_{k}^{N_{c}}\sum_{ij} \sum_{g\in G} \sqrt{h_{\Gamma}} D_{ij}^{\Gamma}(g)\ket{\Gamma,ij }\bra{g}
\end{equation}
Second, if the orbitals are projected into various IRs of the group, then action of $g \in G$ on the $N$-body wave function is 
\begin{equation}
g \mathcal{A}(i^{\Gamma_1} j^{\Gamma_2}...k^{\Gamma_N})=\sum_{\alpha \beta ...\omega} D_{i\alpha}^{\Gamma_1}D_{j\beta}^{\Gamma_2}...D_{k\omega}^{\Gamma_N}\mathcal{A}\left(\alpha^{\Gamma_1} . . . \omega^{\Gamma_N}\right).
\label{eq:g}
\end{equation}
In second quantization, we have $\ket{K}=\prod_i (a_{i\Gamma_i}^\dag)^{K_i}\ket{\Omega}$ transforming under the action of $g\in G$ as 
\begin{equation}
\ket{gK}=\prod_i \left(\sum_m D_{im}^{\Gamma_i}(g)\;a_{m\Gamma_i}^\dag\right)^{K_i}\ket{\Omega}
\end{equation}
The matrix $D^{\Gamma_k}$ is the representation of $g$ in IR $\Gamma_k$.  Finally, we are in a position to utilize the circuit in Figure 1 with inputs $\ket{A}\ket{\Psi}$ where $\Gamma=A$ is the one dimensional trivial representation belonging to all groups where each group element is represented by unity.  
An analysis of this circuit shows the following outcome is obtained:
\begin{eqnarray}
\ket{A}\ket{\Psi}&\mapsto& \frac{1}{\sqrt{h}}\sum_g \ket{g}\ket{\Psi}\mapsto\frac{1}{\sqrt{h}}\sum_g \ket{g} \ket{g\Psi}\\
&\mapsto&\sum_\Gamma\frac{1}{\sqrt{h_\Gamma}} \sum_{ij}\ket{\Gamma, ij}\left( \sum_g \frac{h_\Gamma}{h}D_{ij}^\Gamma(g)^* \ket{g\Psi}\right)\phantom{spc}\\
&=&\sum_\Gamma\frac{1}{\sqrt{h_\Gamma}}\sum_{ij}\ket{\Gamma,ij} \ket{\mathbf{P}_{ij}^\Gamma\Psi} \label{eq:out}
\end{eqnarray}
Measurement of the first register collapses the state into a single IR. While this method is generally applicable, it requires that the irreducible representations are available.  

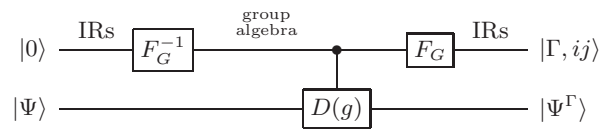
\begin{figure}
\begin{center}
\[\displaystyle\Qcircuit @C=1.5em @R=.75em  {
\lstick{ \ket{0}      }              &  \ustick{\textrm{IRs}}  \qw     &\gate{ F_G^{-1}}     &\qw     & \ustick{\substack{\textrm{group}\\ \textrm{algebra}}}\qw &\ctrl{1}           & \gate{F_G}  &\ustick{\textrm{IRs}}  \qw &\rstick{\ket{\Gamma,ij}}\qw\\
\lstick{\ket{\Psi}} 		& \qw			  &  \qw		   &\qw     & \qw  									      &\gate{D(g)}&\qw&\qw  &\rstick{\ket{\Psi^\Gamma}}\qw \\
}
\]
\end{center}
\caption{The modified phase estimation circuit for the Wigner projection operator.  The input to the readout register is $\ket{0}$ corresponding to $\Gamma=A$, the trivial representation.  The Fourier transform over a group transforms from the space of irreducible representation labels to the group algebra.  The final state is given in \eqref{eq:out} and upon measuring the top register, the state collapses into one of the symmetry adapted states with probability according to the overlap of the initial state with that subspace.  }
\end{figure}

In light of the simple algorithm suggested for first quantized simulations~\cite{Sornborger12}, we point out that the first quantized algorithms for minimal basis molecular hydrogen takes the same number of qubits as the second quantized algorithm.  In the first quantized simulation, each of the two electrons can be in the $1s$ orbital of the left or the right atom requiring only four qubits to store the wave function.  In second quantization, as explained in more detail elsewhere~\cite{Whitfield11}, the simulation also requires four qubits to store the wave function since the spin must be included.  
The spin-free formalism allows the spin in first quantization to be included through the symmetry of the spatial wave function.
Moreover, since all homonuclear diatomic molecules have $D_{\infty h}$ symmetry, the inversion symmetry (which swaps the two nuclei) allows the eigenstates to be uniquely specified.  Details of necessary for the gate decompositions needed for first quantized simulations can be found elsewhere \cite{Kassal08,*Zalka98,*Wiesner96} including the desiderata for fault-tolerant quantum simulations \cite{Jones12}.

As a second example to illustrate the Wigner projections, consider cyclopropenyl cation,  C$_3$H$_3^+$, with $D_{3h}$ symmetry.  We restrict attention to the $C_3$ subgroup which has only one dimensional IRs. As a consequence, the Fourier transform over this group is given by its character table,
\begin{equation}
	F_{C_3}=
{
	\bordermatrix{&\bra{\Gamma_1}&\bra{\Gamma_2}&\bra{\Gamma_3}\cr\\
		\ket{E}&1 & 1   &1   \cr\\
                \ket{C_3}&  1 & e^{2\pi i/3}  &e^{-2\pi i/3}  \cr\\
                \ket{C_3^2}&1   &  e^{-2\pi i/3} & e^{2\pi i/3} }
	}
\end{equation}
Suppose $T(\theta)$ are the phase gates that map $\ket{1_i}\mapsto e^{i\theta}\ket{1_i}$ and $\ket{0_i}\mapsto\ket{0_i}$.  Then in second quantization, the action of $g=C_3$ following \eqref{eq:g} as a circuit is 
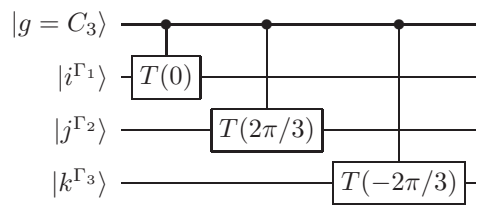
\[
\Qcircuit @C=.4em @R=.4em @!R {
\lstick{ \ket{g=C_3}}                            & \ctrl{1}        &\ctrl{2}            &\ctrl{3}  &\qw \\
\lstick{\ket{i^{\Gamma_1}}} & \gate{T(0)}   &  \qw            &\qw                  & \qw     \\
\lstick{\ket{j^{\Gamma_2}}  }& \qw               &\gate{T(2\pi/3)}          &\qw                   & \qw     \\
\lstick{\ket{k^{\Gamma_3}} }& \qw               &\qw              &\gate{T(-2\pi/3)}             & \qw     \\
}
\]
The inverse of this circuit gives the action of $C_3^2$ and the identity operation acts trivially.  A full example in second quantization would require a qutrit for the readout register and 6 qubits to simulate the H\"uckel model where each qubits corresponds to one of three $2p_z$ paired with either spin function $\alpha$ or $\beta$.

The modifications to the Abrams-Lloyd algorithm enables spin-free quantum simulation thereby lowering the spatial requirements without requiring any other modifications of the underlying algorithms.  Unfortunately, in the second quantized setting the results are not as dramatic.  The preparation of multi-determinants states using the techniques in Ref.~\cite{Ortiz01} will likely be more efficient for preparing spin eigenstates in second quantization than the Wigner projection operator techniques.

In the second quantized algorithm, the number of operations needed to simulate the evolution of the Hamiltonian is independent of $N$ since all Fock subspaces $F(k,M)$ with $k\leq N$ particles are simulated simultaneously \cite{Whitfield11,Seeley12}.  This leads to a significantly greater number of operations at asymptotically large system sizes \cite{Abrams97}.  If one could implement the Hamiltonian in a particular Fock sector, this would provide a great advantage.  While we have focused on the permutation group to enable spin free simulations in the first quantized simulations, in second quantization, the unitary group, generated by $E_{ij}=\sum_{\sigma=\alpha}^\beta a_{i\sigma}^\dag a_{j\sigma}$, will likely provide new strategies for simulation. 

Let us close the article by reiterating the importance of translating existing techniques and heuristics to the quantum domain as exemplified by the present work. While many experimental groups worldwide are pushing for realizations of quantum computers, it is important that the theoretical advancements strive to yield feasible simulations that enable quantum computation to produce results relevant to the wider community of scientists and engineers.  Analog quantum simulation has advanced to a nearly competitive stance, however digital quantum simulation is still in early development~\cite{NatureInsights}.  The question of when quantum computers will provide a competitive alternative is entirely moot until a test set of simulations, e.g.~the G1 molecular test set~\cite{Curtiss90}, have been performed experimentally~\footnote{The larger molecules in the G1 set have on the order of 30 electrons. For example, molecules Si$_2$H$_6$ and SO$_2$ have 34 and 32 electrons which, in the cc-pVTZ basis, require 152 and 94 spatial orbitals, respectively.}.  Only then can the accuracy of quantum simulations be compared against the full configuration interaction method it is meant to displace.

{\bf Acknowledgments}:  The author thanks the Vienna Center for Quantum Science and Technology Postdoctoral Fellowship award for support.  Useful discussions with M. B. \c{S}ahino\v{g}lu, N. Wiebe, M. R\"otteler, S. Jordan and J. Taylor were appreciated.

\end{document}